\documentclass[mypaper,7pt,twoside]{CoAst}
\usepackage{epsf,graphicx,fancyhdr,url}
\renewcommand{\chaptermark}[1]{\markboth{#1}{}}

\renewcommand{\headrulewidth}{0pt}
\newcommand{\kopf}{\small\itshape Comm. in Asteroseismology \\ Vol. number, publication date (will be inserted in the production process)}

\newcommand{\Authors}[1]{\begin{center}\normalsize\bf\sf #1 \end{center}}

\renewcommand{\author}[1]{\begin{center}\normalsize\bf\sf #1 \end{center}}
\newcommand{\Address}[1]{\begin{center}\small\sf #1 \end{center}}

\newcommand{\Objects}[1]{{\vspace{2mm}\small \noindent  \hspace*{0mm}Individual Objects: } \small #1 \normalsize}

\renewenvironment{abstract}{\section*{Abstract}\normalsize\sf}{}
\newcommand{\References}[1]{\begin{flushleft}{\large References\\}\vspace*{2mm}\small #1 \end{flushleft}}

\newcommand{\chapterCoAst}[2]{\chapter[\sf\normalsize #1\\ \footnotesize \hspace*{5mm}by #2 \sf\normalsize][]{#1\\}\rhead[\fancyplain{}{\sf\footnotesize \center{#1}}]{\fancyplain{}{\sffamily\thepage}}\lhead[\fancyplain{\kopf}{\sffamily\thepage}]{\fancyplain{\kopf}{\sf\footnotesize \center{#2}}}}

\renewcommand{\chaptername}{}

\newcommand{\acknowledgments}[1]{\vspace*{5mm}\noindent  \textbf{Acknowledgments.} #1}

\def\rfr{\smallskip\par\noindent
        \hangindent=7truemm
        \hangafter=1}

\def\hr{\hbox{$^{\rm h}$}}                 
\def\fdeg{\hbox{$.\!\!^\circ$}}            
\def\fsec{\hbox{$.\!\!^{\rm s}$}}          
\def\arcm{\hbox{$^\prime$}}                
\def\farcs{\hbox{$.\!\!^{\prime\prime}$}}  
\def\day{\hbox{$^{\rm d}$}}                
\def\fday{\hbox{$.\!\!^{\rm d}$}}          
\def\mm{\hbox{$^{\rm m}$}}                 
\def\fmm{\hbox{$.\!\!^{\rm m}$}}           

\usepackage{subfig}  
\usepackage{amstext} 

\begin{document}
\sf

\chapterCoAst{A new eclipsing binary system with a pulsating component detected by CoRoT$^\ast$}
{K.\,Sokolovsky, C.\,Maceroni, M.\,Hareter, C.\,Damiani, L.\,Ba\-la\-guer-\-N\'u\-\~nez, and I.\,Ribas} 
\Authors{K.\,Sokolovsky$^{1,2}$, C.\,Maceroni$^3$, M.\,Hareter$^4$,
C.\,Damiani$^3$, L.\,Ba\-la\-guer-\-N\'u\-\~nez$^5$, and I.\,Ribas$^6$} 
\Address{$^1$ Max-Planck-Institute f\"ur Radioastronomie, Auf dem H\"ugel 69, D-53121 Bonn, Germany\\
$^2$ Astro Space Center of Lebedev Phys. Inst., Profsoyuznaya 84/32, 117997 Moscow, Russia\\
$^3$ INAF - Osservatorio Astronomico di Roma, via Frascati 33, Monteporzio C.,  Italy\\
$^4$ Institut f\"ur Astronomie, Universit\"at Wien, T\"urkenschanzstra{\ss}e 17, 1180 Vienna, Austria\\
$^5$ Departament d'Astronomia i Meteorologia-ICC-IEEC, Universitat de Barcelona, Av. Diagonal, 647, 08028 Barcelona, Spain\\
$^6$ Institut de Ci\`encies de l'Espai (CSIC-IEEC), Campus UAB, Facultat de Ci\`encies, Torre C5, parell, 2a pl., E-08193 Bellaterra, Spain
}

\footnotetext[1]{The CoRoT space mission was developed and is operated
by the French space agency CNES, with participation of ESA's RSSD and
Science Programs, Austria, Belgium, Brazil, Germany and Spain.}

\noindent
\begin{abstract}
We report the discovery of CoRoT~102980178
($\alpha = 06\hr50\mm12\fsec10$, $\delta = -02^\circ41\arcm21\farcs8$, J2000) 
an Algol--type eclipsing binary system with a pulsating component (oEA).
It was identified using a publicly available 55 day long monochromatic
lightcurve from the CoRoT initial run dataset (exoplanet field).
Eleven consecutive $1\fmm26$ deep total primary and the equal number of $0\fmm25$ deep secondary 
eclipses (at phase $0.50$) were observed. The following light elements for the primary eclipse
were derived:
$\textit{HJD}_{\text{MinI}} = 2454139.0680 + 5\fday0548 \times \textit{E}$.
The lightcurve modeling leads to a semidetached configuration with the
photometric mass ratio $q=0.2$ and orbital inclination $i = 85^\circ$.
The out-of-eclipse lightcurve shows ellipsoidal variability and positive
O'Connell effect as well as clear $0\fmm01$ pulsations with the dominating
frequency of $2.75$~c/d.
The pulsations disappear during the primary eclipses, 
which indicates the primary (more massive) component to be the pulsating star.
Careful frequency analysis reveals the second independent pulsation frequency of
$0.21$~c/d and numerous combinations of these frequencies with the binary
orbital frequency and its harmonics. 
On the basis of the CoRoT lightcurve and ground based multicolor photometry, we favor 
classification of the pulsating component as a $\gamma$~Doradus
type variable, however, classification as an SPB star cannot be excluded. 
\end{abstract}

\Objects{CoRoT~102980178} 

\section*{Introduction}

Pulsating stars in eclipsing binary systems are objects of considerable
astrophysical interest. It is well known that in double--line
binaries masses, radii and luminosities of the components may be directly 
determined which allows one to constrain pulsation models.
Possible influence of one component of a binary system on 
pulsations of the other component (through mass transfer or
tidal interaction) is an interesting and still largely unexplored topic.
A case of tidal excitation of $\gamma$~Doradus type pulsations was reported
by Handler et al. (2002; see also references to some theoretical
discussions therein). Finally, binary
systems with oscillating component offer an intriguing possibility of direct observations of radius
change during the pulsation cycle as periodically changing moments of 1st,
2nd, 3rd and 4th contact (in the case of radial pulsations), and in
general case -- possibility of pulsation mode identification by use of 
the secondary component as a spatial filter (B{\'{\i}}r{\'o} 
\& Nuspl 2005). Altogether, identification of pulsating stars of different
types in eclipsing systems is important.

The combination of pulsations and eclipses is often found among the evolved stars like 
symbiotic binaries (e.g. Chochol \& Pribulla 2000) and cataclysmic variables
(e.g. Araujo-Betancor et al. 2005). Four Type~I or
II Cepheids in eclipsing binaries have recently been identified
(see Antipin et al. 2007 and references therein), 
the confirmation of one more possible object of this type is still pending
(Khruslov 2008). The discovery of a red giant with solar-like
oscillations in a long-period eclipsing binary system was recently announced
by Hekker et al. (2010).

Semi-detached eclipsing binaries with a pulsating primary component which is
close to the main sequence were dubbed ``oscillating Algols'' 
(oEA, Mkrtichian et al. 2002). About 
twenty such systems are known to date (Mkrtichian et al. 2007) most
of them containing $\delta$~Scuti type components. The first eclipsing
binary containing $\gamma$~Doradus type pulsating star (VZ~CVn) was
reported by Ibano{\v g}lu et al. (2007). Recently 
Damiani et al. (2010) and Maceroni et al. (2010)
identified two $\gamma$~Doradus candidates in eclipsing binaries using 
CoRoT photometry (CoRoT~102931335 and CoRoT~102918586).

CoRoT (Convection, Rotation and planetary Transit; Fridlund et al 2006) is a space experiment
devoted to study of convection and rotation of stars and to
detection of planetary transits. 
CoRoT data become publicly available one year after release to the Co--Is of the mission from 
the CoRoT archive: \url{http://idoc-corot.ias.u-psud.fr/}

In this paper we announce the discovery of a new oEA system identified using publicly 
available data from the
CoRoT initial run (exoplanet field). An almost uninterrupted 55 day long monochromatic 
CoRoT lightcurve together with ground--based multicolor photometry allow us to identify 
the pulsating component as a likely $\gamma$~Dor type variable, however
classification as a Slowly Pulsating B (SPB) star can not be ruled out. The
object is an interesting target for a spectroscopic follow--up.

\section*{Observational data}

\subsection*{CoRoT photometry}

\begin{figure}
  \centering
  \includegraphics[width=0.7\textwidth,angle=0]{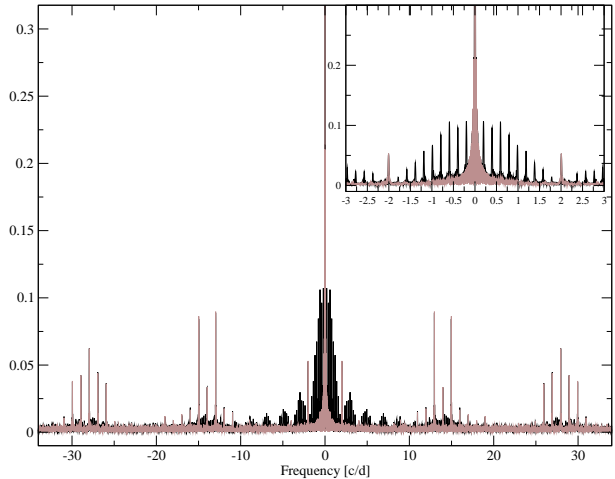}
  \caption{Spectral windows for the outlier-rejected data (light) and for the
data excluding the primary eclipses (dark), which were used for frequency analysis.
The inset shows a zoom-in to the lower frequencies.}
  \label{fig:Spw}
\end{figure}

CoRoT~102980178 (USNO-B1.0~0873-0161681, coordinates:
$\alpha = 06\hr50\mm12\fsec10$, $\delta = -02^\circ41\arcm21\farcs8$, J2000, Monet et al. 2003)
was identified by us as an Algol type eclipsing binary with oscillating component
(oEA star) after visual inspection of the CoRoT lightcurve.
The star was also independently identified as an eclipsing binary by
Debosscher et al. (2009) and Carpano et al. (2009).
The CoRoT lightcurve covers almost 55 days from JD~$2454138.1$ to JD~$2454192.8$
with $512$~sec time sampling. We have removed data points with large error--bars
which were typically upward outliers caused by high--energy particles hitting
the CCD detector inside the measurement aperture. Most of these events
occur when the satellite crosses the South Atlantic Anomaly (SAA). Elimination of
these data points introduces small gaps in the lightcurve, with some of
these gaps occurring quasi--periodically (due to SAA crossing). However,
the number of discarded data points is relatively small ($1168$ out of
$9229$, $12.6\%$). To characterize the effect of periodic data rejection on
the frequency analysis we have constructed the spectral window
plot presented on Fig.~\ref{fig:Spw}. The amplitudes of the alias
frequencies are of the order of 10~\% or below.
We have also corrected the lightcurve for a small
($0.211$~mmag/day) downward trend which is probably a result 
of gradual drift of the satellite pointing or some kind of instrumental decay. 
A detailed discussion of noise properties of the CoRoT data may be found in
Aigrain et al. (2009) and Auvergne et al. (2009).

\subsection*{Ground--based multicolor data}

In the framework of ground support to the CoRoT mission, we obtained 
Str{\"o}mgren $uvby$ photometry of this object
using the Wide Field Camera (WFC) on the 2.5~m Isaac Newton Telescope (INT) Roque Muchachos
Observatory, La Palma, Canary Islands. The observations were conducted on
JD~$2454170.391$ which corresponds to the orbital phase $0.197$ of the binary system.
The object was also observed in Johnson $B$ and $V$ filters by
Deleuil et al. (2006) using the same telescope and camera.
Infrared $JHK_s$ photometry of this object is available
from the 2MASS catalog (Skrutskie et al. 2006). The 2MASS
observations were conducted on JD~$2451526.704$ which corresponds to the
orbital phase $0.191$ (by a lucky coincidence it is very close to the phase of our
Str{\"o}mgren photometry). The great advantage of the 2MASS photometry is that it
was obtained using a camera which utilizes a beam splitter producing truly simultaneous
measurements in three filters. Therefore, 2MASS colors are not
distorted by stellar variability.

All available color information is summarized in Table~\ref{tab:color}.

\begin{table}[ht]
  \centering
  \caption{Multicolor photometry of CoRoT~102980178}
  \begin{tabular}{ r r l c }
Parameter  & Value    &  Error      &  Origin  \\
  \hline
  $b-y = $ & $0.598 $ & $ \pm0.032$ & INT/WFC  \\
  $  y = $ & $15.723$ & $ \pm0.034$ & INT/WFC  \\
  $m_1 = $ & $-0.042$ & $ \pm0.11$  & INT/WFC  \\
  $c_1 = $ & $0.891 $ & $ \pm0.086$ & INT/WFC  \\
  $ B = $  & $16.58~ $ & $ \pm0.45$  & INT/WFC  \\
  $ V = $  & $15.73~ $ & $ \pm0.19$  & INT/WFC  \\
  $ J = $  & $13.411$ & $ \pm0.028$ &  2MASS   \\
  $ H = $  & $12.759$ & $ \pm0.023$ &  2MASS   \\
  $K_s = $ & $12.528$ & $ \pm0.027$ &  2MASS   \\
  \hline
  \end{tabular}
  \label{tab:color}
\end{table}

\subsection*{Interstellar extinction estimation}

In the absence of spectroscopic data it is important to constrain the
influence of interstellar reddening, because information about intrinsic
colors is necessary for unambiguous classification of the pulsating star.

According to the standard tables by Schlegel et al. (1998)
the {\em total} Galactic extinction in the direction of the object is large:
$A_V = 3\fmm860$, $E(B-V) = 1\fmm164$. The Schlegel et al. estimation is based on
direct observations of far--infrared emission of interstellar dust, the material which causes
absorption in visible and near--infrared bands. However, as it is noted by the authors, the
extinction estimation for this position may be unreliable because of the low Galactic latitude
($b = -1\fdeg50 $). There is an alternative way to estimate
the Galactic extinction which is based on the relation between HI
column density ($N_{HI}$) and extinction caused by dust:
$N_{HI}/A_V = 1.79 \times 10^{21}$~cm$^{-2}$~mag$^{-1}$
(Predehl \& Schmitt 1995). We estimate $N_{HI}$ in the direction of
the object using $21$~cm radio observations from the Leiden/Argentine/Bonn Galactic
HI Survey (Kalberla et al. 2005, see also
\url{http://www.astro.uni-bonn.de/~webaiub/english/tools_labsurvey.php}).
From these data we obtain $N_{HI} = 0.682 \times 10^{22}$~cm$^{-2}$
corresponding to $A_{V}(HI) = 3\fmm810$ which is in good agreement with
the estimation obtained using the Schlegel et al. tables.

Unfortunately, the distance to the binary system is unknown, as well as
the exact distribution of the absorbing material along the line of sight.
The intrinsic color of the system lies somewhere between the observed color
and the color corrected for the total Galactic reddening along the line of
sight. However, even reliable upper limits on the Galactic reddening will be
useful for the following discussion.

\section*{Interpretation and Modeling}

\begin{figure}
  \centering
  \includegraphics[width=0.8\textwidth,angle=0]{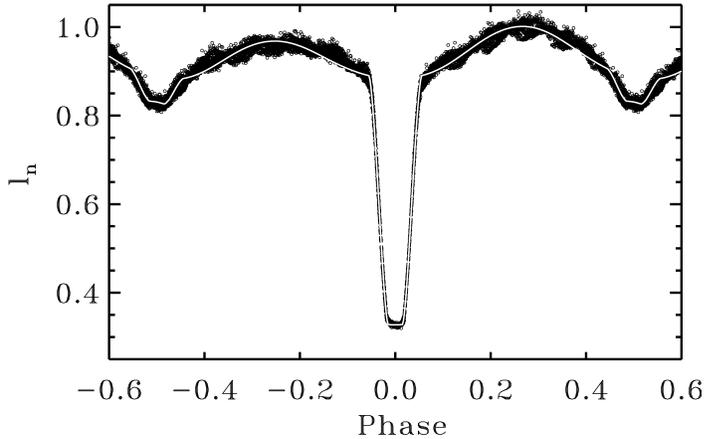}
  \caption{Detrended lightcurve of CoRoT~102980178 phased with light
  elements $\textit{HJD}_{\text{MinI}} = 2454139.0680 + 5\fday0548 \times  \textit{E}$.
  The solid line represents the F-primary model lightcurve.}
  \label{fig:phase_lc}
\end{figure}

The lightcurve of CoRoT~102980178 shows eleven $1\fmm26$ deep primary
and the equal number of $0\fmm25$ deep secondary eclipses (at phase $0.50$).
The nearly flat primary minimum bottom indicates that the eclipses are total. Ellipsoidal
variability is evident in the out of eclipse lightcurve
(Fig.~\ref{fig:phase_lc}).
The Maximum preceding the primary minimum is $0\fmm04$ fainter then the maximum
which follows it (positive O'Connell effect, see Davidge \& Milone 1984
and Liu \& Yang 2003 for a discussion of the effect).
The period analysis with the Lafler \& Kinman (1965) method leads to the
following light elements:
$$\begin{array}{r r c r c c}
\textit{HJD}_{\text{MinI}} = & 2454139.0680 & + & 5\fday0548    & \times & \textit{E} \\  
                           &    \pm0.0007 &   & \pm0\fday0170 &        &   \\
\end{array}$$
Individual minima times were estimated using Kwee \& van Woerden (1956)
method and combined to produce a single primary minimum epoch.
Its $1 \sigma$ uncertainty was estimated from scatter of individual
measurements on the Observed minus Calculated ($O-C$) plot produced with
the above light elements.
Period uncertainty was estimated following Schwarzenberg-Czerny (1991).

\begin{figure}
  \centering
  \includegraphics[width=0.9\textwidth]{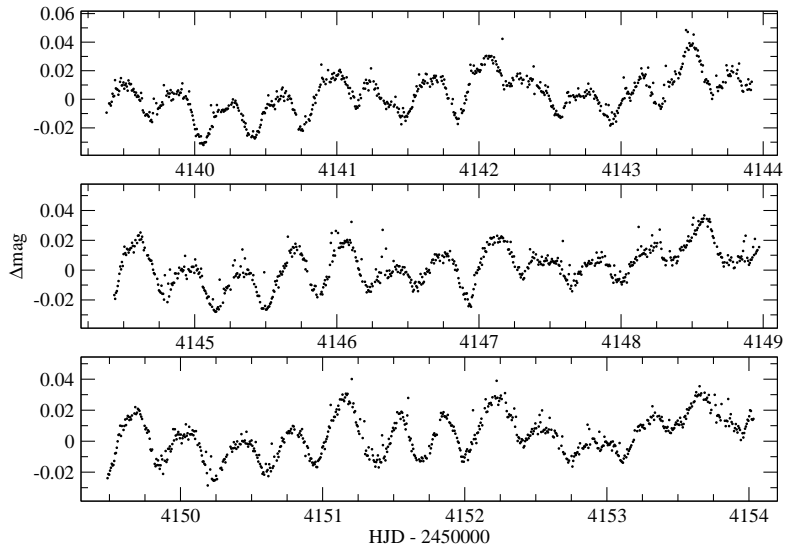}
  \caption{The residual lightcurve (F-primary model).}
  \label{fig:res}
\end{figure}

\begin{figure}
  \centering
  \includegraphics[width=1.0\textwidth]{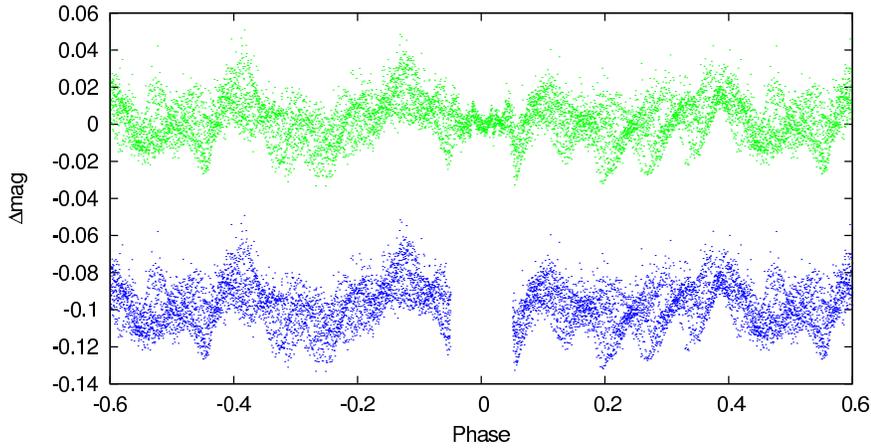}
  \caption{The residual lightcurve (F-primary model) folded with the binary orbital
  period. The lower curve represents the section of residual lightcurve used
  for frequency analysis, it was shifted $0\fmm1$ along the vertical axis for
  visibility.}
  \label{fig:res_orb}
\end{figure}

Superimposed on this classical Algol--type lightcurve are $0\fmm01$
oscillations with the period of $0\fday36372$ (see Fig.~\ref{fig:res} and the detailed discussion below). 
The oscillations are evident at all phases except during the primary minimum (Fig.~\ref{fig:res_orb}).


%

\subsection*{Ruling out an optical blend}

\begin{figure}
  \centering
  \subfloat[CoRoT image]{\label{fig:corot_image}\includegraphics[width=0.5\textwidth]{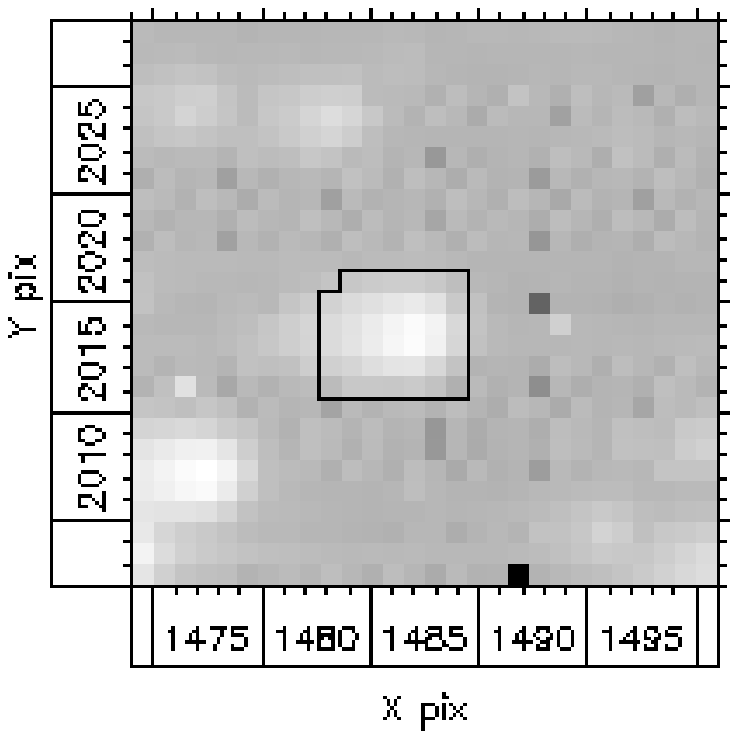}}                
  \subfloat[DSS finding chart]{\label{fig:dss_finding_chart}\includegraphics[width=0.5\textwidth]{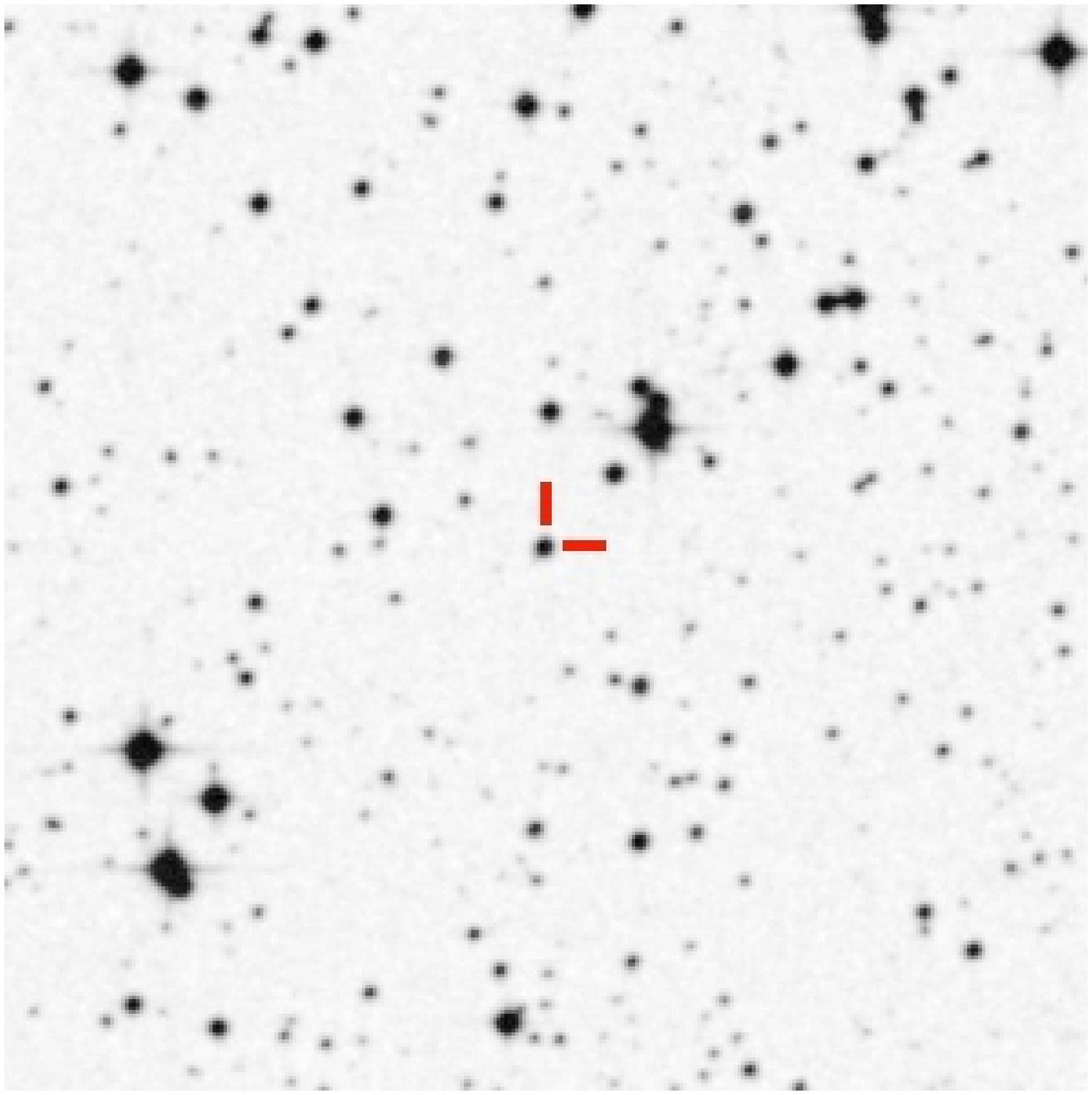}}
  \caption{(a) CoRoT image of the variable star, marked area is the
  measurement aperture. The image is in detector coordinates, east is up, north is
  to the left. (b) 5\arcm~x~5\arcm~ DSS2 red
  image centered on CoRoT~102980178 (north is up, east is to the
  left).}
  \label{fig:charts}
\end{figure}

When analyzing a variable star showing two different types of variability 
(e.g. eclipses and pulsations) it is possible that the observed 
variability comes from two unrelated objects which just happened to 
be on the same line of sight. For the variable star described here, 
this possibility may be ruled out due to the following reasons:
1) inspection of both the original CoRoT image (Fig.~\ref{fig:corot_image})
and the DSS image (Fig.~\ref{fig:dss_finding_chart}) reveal no other stars
in the immediate vicinity of the variable and 2) pulsations disappear during
the primary eclipse, which proves that the pulsating star is actually blocked from view.

\subsection*{Oscillating component classification}

\begin{figure}
  \centering
  \includegraphics[width=0.8\textwidth]{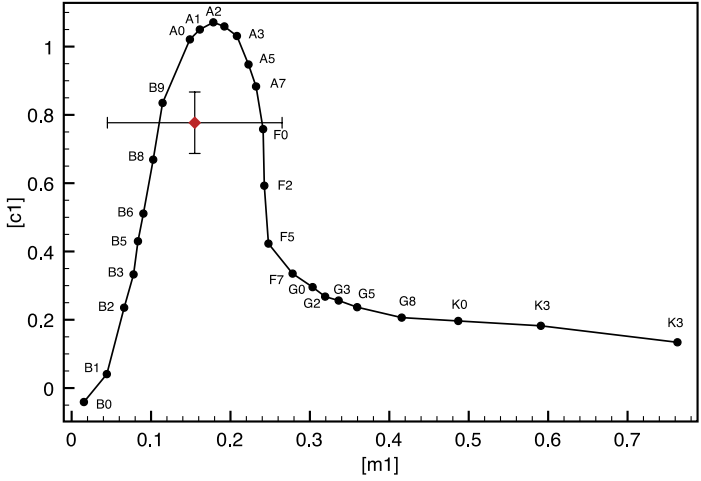}
  \caption{$[c_1]$---$[m_1]$ plot from Moon (1986) with position of CoRoT~102980178
  and its $1\sigma$ uncertainty indicated.}
  \label{fig:moon}
\end{figure}

Following Moon (1986) we calculate reddening--free indices
$$[c_1]=c_1-0.19(b-y) = 0.777 \pm0.087$$
$$[m_1]=m_1+0.33(b-y) = 0.155 \pm0.111$$
which are consistent with a main sequence star of spectral class B$8$-$9$ or
A$7$-F$1$ within the uncertainty (see Fig.~\ref{fig:moon}).

The Johnson $B$ and $V$ band observations with INT/WFC by Deleuil et al. (2006)
(see Table.~\ref{tab:color})
have large error bars and where not used in the
analysis, however, we note that the $(B-V)$ color of the star do not
contradict our proposed spectral classification after accounting for the
uncertainties in the Galactic extinction.

2MASS infrared colors can also be used to estimate spectral class.
Accounting for the uncertainty in the Galactic reddening discussed above, 
the infrared colors correspond to spectral class in the range early A -- early M
(Bessell \& Brett 1988). 
However, it is possible that the secondary component of the
binary system dominates the infrared light, so this spectral type estimation may
correspond to the secondary star instead of the pulsating component.

The multicolor photometry excludes early B spectral class, but is consistent
with a reddened late B or F class star. 

The period and amplitude of oscillations are typical for $\gamma$~Doradus
type but also within the possible range for young Slowly Pulsating B (SPB) 
and $\beta$~Cephei stars.
Information about the spectral class is crucial to distinguish between these
possibilities: $\beta$~Cephei variables are early--type B stars (B$0$ --
B$2.5$), SPB stars have spectral types between B$2$ and B$9$ (Stankov \& Handler 2005)
while $\gamma$~Doradus variables
are typically early F type stars (e.g. Rodr{\'{\i}}guez 2002,
Henry et al. 2007). $\beta$~Cephei classification can be excluded on the
basis of multicolor photometry. Note also, that according to
Pigulski \& Pojma{\'n}ski 2009, $\beta$~Cephei variables are
characterized by periods $<0\fday35$ while the GCVS (Samus et al. 2009)
gives $0\fday6$ as the maximum period for this variability class.
SPB classification is consistent with the Str{\"o}mgren photometry, but SPB
stars are usually characterized by longer pulsation periods of $0\fday5$ --
$5\day$ (Thoul 2009) and therefore, we consider classification of
the oscillating component as an SPB star to be less likely.
$\gamma$~Doradus stars are characterized by periods ranging from $0\fday4$ to $3\day$ (Kaye et al 1999).
Therefore, $\gamma$~Doradus type variable is the most probable
classification of the oscillating component.

\subsection*{Binary system modeling}

For the construction of the binary system model we have cleaned the lightcurve
from outlier points and corrected it for the linear systematic trend. No
prewhitening for the pulsations was done. The oscillations should not affect the
binary model because they are significantly smaller that the main features of the
eclipsing lightcurve.

The light curve was analyzed with the {\em PHOEBE} software
(Pr{\v s}a \& Zwitter 2005, see also \url{http://phoebe.fiz.uni-lj.si}~), which provides 
a user-friendly interface and additional capabilities to the 
Wilson \& Devinney (1971) code.
The software was recently modified to take into account CoRoT transmission functions, both for
flux and limb darkening computations (for details see Maceroni et al. 2009).
Being the orbit-unrelated variability  of much smaller amplitude and shorter period than the eclipses,
we just fitted the light curve folded in phase according to the ephemeris presented above
(Fig.~\ref{fig:phase_lc}).
 
 After an initial screening of the parameter space, we proceeded to the light curve solution,
using -- at different stages --  both the differential correction and the Nelder and Mead's  Simplex algorithm.
The fit adjustable parameters were: the inclination, $i$, the mass ratio, $q=M_2/M_1$, the
surface potentials $\Omega_{1,2}$ (which, together with q, determine the components fractional radii $r_{1,2}$), 
and the secondary effective temperature, $T_{\rm eff,2}$.
The albedo  $a_i$ and gravity darkening coefficients $\beta_i$ were fixed at the
theoretical values ($a_{1,2} = 1.0, 0.5$ and $\beta_{1,2} = 1.0, 0.32$, respectively, for stars with radiative or convective envelopes). 
We adopted a non-linear limb darkening law, whose coefficients were interpolated from the {\em PHOEBE} tables on-the-fly, 
according to the star temperature and surface gravity ($\log g$). For the F-primary model (see below) we chose a logarithmic limb darkening law,
for the B-primary one we chose a square-root law.
The primary luminosity in the CoRoT passband, $L_1$, was computed at each iteration  in {\em PHOEBE}, rather 
than adjusted, to enhance convergence ($L_2$ is derived from the other parameters and  model atmospheres). 

We started from a detached configuration but in all cases the
iterative solution evolved towards a semi-detached configuration (with the less massive star
in contact). We checked, as well, the effect of non-zero eccentricity, which however did not
provide any improvement of the fit.

It is well known that the solution of a single passband light curve is mainly sensitive
to the $T_{\rm eff}$ relative values, therefore a shift in $T_{\rm eff,1}$ yields  a  shift in 
the same direction of $T_{\rm eff,2}$.  
It was therefore possible to fit the light curve incorporating different assumptions about the
temperature of primary (pulsating) component: $T_{\text{eff}~1}=7000~K$ (early F spectral
class) and $T_{\text{eff}~1}=11400~K$ (late B class). As expected, we obtained
 different temperatures of the secondary: $T_{\text{eff}~2}=4692$
(K class) and $T_{\text{eff}~2}=6162$ (late F) for the first and
second model respectively. Apart from that, the models are very similar.
The uncertainty of $T_{\text{eff}~2}$ is dominated by our poor knowledge of $T_{\text{eff}~1}$.
For a particular assumed value of $T_{\text{eff}~1}$ the formal error bars of
$T_{\text{eff}~2}$ calculated from the least square fit are less than $10$~K.

To achieve an acceptable light curve solution the non-negligible O'Connell effect shall also
be modeled, and for this purpose we introduced a bright spot on the
primary component. Its parameters (location, size and temperature contrast factor,
$T_{\text{spot}~1}/T_{\text{eff}~1}$)
where not adjusted, but determined by trial and error, by putting 
the spot on the stellar equator and varying only the spot longitude and contrast factor  
(as the size  and the latitude are highly correlated with them, see for
example the discussion by Maceroni \& van't Veer 1993).
It is reasonable to expect the existence of such a spot as
the result of mass accretion from the secondary component.  However different solutions
(such as a dark spot on the colder secondary star, due in this case to surface
activity), cannot be ruled out.
The model parameters are summarized in Table~\ref{tab:model}.

\begin{table}[ht]
 \begin{centering}
  \caption{Model parameters of the binary system}
  \begin{tabular}{ r c c }
Parameter  & Model with F primary  &  Model with B primary  \\
  \hline
  assumed $T_{\text{eff}~1}$   & $7000~K$ & $11400~K$ \\
 $T_{\text{eff}~2}$     & $4692~K$ & $6162~K$  \\ 
 $q=M_{2}/M_{1}$    & $0.206 \pm 0.001$  & $0.184 \pm 0.001 $ \\
  $i$                             & $84\fdeg61 \pm 0.05$ & $85\fdeg61 \pm 0.04$ \\
$r_1$                          & 0.1068 $\pm$ 0.0002 & 0.1064 $\pm$ 0.0002   \\
$r_2$                          & 0.2447 $\pm$ 0.0003 & 0.2363 $\pm$ 0.0003\\
$L_2/L_1   $              & 0.7070 $\pm$ 0.0005 &  0.7183 $\pm$0.005\\
$T_{\text{spot}~1}/T_{\text{eff}~1}$ & $1.22$ & $1.50$ \\
 
  \hline
  \end{tabular}
  \label{tab:model}
  \end{centering}

 The spot angular radius was fixed to  15$^{\circ}$ its colatitude to 90, the 
longitude of 240$^{\circ}$ and the temperature contrast factors in the table
were found by trial and error. The uncertainties are formal fit errors ($1\sigma$).

\end{table}

One of the lightcurve features which are not reproduced by the models is a
small difference in depth among primary eclipses. 
The eleven observed eclipses are not sufficient to draw a conclusion about
possible periodicity of the primary eclipse depth variation. The models also predict
symmetric lightcurve shape during the total eclipse phase which is not
observed. These two effects may be caused by starspots on the secondary
component.

As can be seen from the Fig.~\ref{fig:phase_lc}, there are some residual systematic
deviations of the model with respect to observations. This will imply the
presence of the orbital frequency ($\text{f}_{\text{orb}}$) and its multiples 
(typically the even ones) in the residual
lightcurve spectrum which is discussed in the following section.

\subsection*{Pulsation frequency analysis}

The residual lightcurve resulting from subtraction of the binary model
(version with early-F type primary) is presented on Fig.~\ref{fig:res}.
It shows clear pulsation with a dominant frequency $\sim2.75$~c/d.
If folded with the binary orbital period (see Fig.~\ref{fig:res_orb}) the
residual lightcurve shows a rather
complicated pattern instead of uncorrelated data. This means that the main
pulsation frequency is close to an integer multiple of the binary orbital 
frequency ($\text{f1}/\text{f}_{\text{orb}} = 13.898 \sim 14$).
The pulsation disappears during the primary eclipse (phases $-0.05$ -- $0.05$ on
Fig.~\ref{fig:res_orb}), and this phase interval was excluded from
the frequency analysis. In order to check for possible artifacts introduced
due to omitting parts of the lightcurve, the analysis was repeated including
observations during total eclipses. The introduced artifacts were 
found to be negligible.

\begin{figure}
  \centering
  \includegraphics[width=0.7\textwidth]{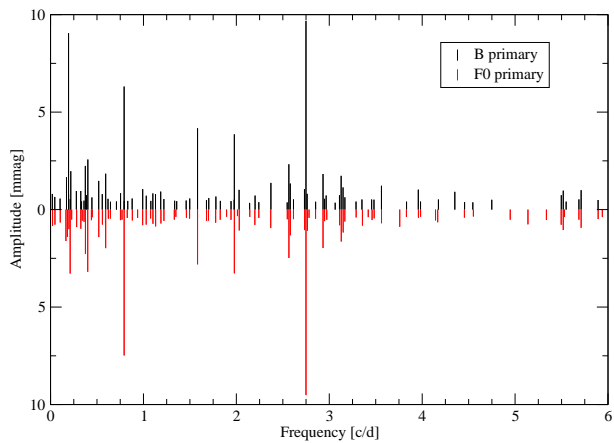}
  \caption{Amplitude spectrum of the residual lightcurve after subtracting
  models with B (black) and F (red, plotted downwards for better
  visibility) primary.}
  \label{fig:Freqvergleich}
\end{figure}

The frequency analysis was performed using the {\em SigSpec} software
(Reegen 2007), leading to 99
formally significant frequencies including the CoRoT orbit and its
harmonics. The first few frequencies are listed in the
Table~\ref{tab:freq}.
The search for combination frequencies was performed using the software
{\em COMBINE} developed by P.~Reegen (see \url{http://www.SigSpec.org}).  The frequency
errors were estimated following Kallinger et al. 2008.
The orbital frequency of the
binary system was added to the frequency list, because from the plot of the
resulting frequencies (Fig.~\ref{fig:Freqvergleich}) it is obvious that there are many combinations with
the orbital frequency. 

Only two frequencies (f1 and f3 in
Table~\ref{tab:freq}) appear to be genuine, which means not explained by a
combination; f3 is close to $\text{f}_\text{orb}$ but the frequency
errors are small enough to conclude that the frequencies are significantly
different (at $10 \sigma$ level).

\begin{table}[ht]
 \centering
 \caption{Frequencies detected in the residual lightcurve}
 \begin{tabular}{ r c c | c c }
   & \multicolumn{2}{c}{Model with F primary} & \multicolumn{2}{c}{Model with B primary} \\
   &        Frequency      &   Amplitude   &     Frequency     &   Amplitude   \\
   &          [c/d]        &    [mmag]     &        [c/d]      &    [mmag]     \\
 \hline
                            f1:  & $2.7494\pm0.0008$ & $9.52\pm0.42$ & $2.7494\pm0.0009$ & $9.68\pm0.48$ \\
 ($4 \text{f}_\text{orb}$)  f2:  & $0.7917\pm0.0008$ & $7.48\pm0.35$ & $0.7920\pm0.0009$ & $6.31\pm0.32$ \\
                            f3:  & $0.2115\pm0.0012$ & $3.28\pm0.25$ & $0.2181\pm0.0021$ & $1.97\pm0.23$ \\
 ($2 \text{f}_\text{orb}$)  f4:  & $0.4004\pm0.0015$ & $3.19\pm0.26$ & $0.4012\pm0.0016$ & $2.57\pm0.22$ \\
($10 \text{f}_\text{orb}$)  f5:  & $1.9784\pm0.0017$ & $3.27\pm0.30$ & $1.9783\pm0.0014$ & $3.86\pm0.30$ \\
 ($8 \text{f}_\text{orb}$)  f6:  & $1.5823\pm0.0017$ & $2.81\pm0.27$ & $1.5822\pm0.0013$ & $4.17\pm0.29$ \\
($13 \text{f}_\text{orb}$)  f7:  & $2.5657\pm0.0019$ & $2.48\pm0.25$ & $2.5650\pm0.0019$ & $2.32\pm0.25$ \\
   ($\text{f}_\text{orb}$)  f8:  & $0.1978\pm0.0031$ & $1.00\pm0.17$ & $0.1940\pm0.0008$ & $9.05\pm0.40$ \\
       & \multicolumn{2}{c}{$\dots$}       & \multicolumn{2}{c}{$\dots$}       \\
 \hline
 \end{tabular}
 \label{tab:freq}
\end{table}

Analysis of the residual lightcurve after subtraction of the B primary model
leads to similar results (Table~\ref{tab:freq}). The one major difference
is that f8 becomes second most powerful harmonic in the
spectrum. Its amplitude is nine times larger compared to residuals from the
binary model with early-F type primary. 
f8 in the B primary model residual spectrum can be associated 
with the orbital frequency ($\text{f}_\text{orb}=0.1978 \pm0.0007$),
the difference between them is less then $4 \sigma$. 
Most probably, it is not a real frequency but
an artifact introduced by imperfect binary model subtraction.





All the other frequencies can be explained by combinations mainly with the
orbital frequency. This is what can be expected from a close binary because
the primary component is likely to be distorted by tidal forces.
If the F type primary hypothesis is accepted, the solution for the
independent frequencies has two components: f1 and f3.

\section*{Conclusions}

We have identified CoRoT~102980178 as a semidetached eclipsing
binary with a $5\fday0548$ orbital period.
It shows an Algol--type lightcurve with $1\fmm26$ deep primary
and $0\fmm25$ deep secondary eclipses (in CoRoT photometric band).
 
Our lightcurve modeling leads to a semidetached configuration with the
photometric mass ratio $q=0.2$ and orbital inclination $i = 85^\circ$.
We note, that the inverse problem of binary model fitting
using only a broad passband light curve is degenerate and the proposed
solution might not be unique.

The more massive component of the system is pulsating with the primary
frequency of $2.7494$~c/d ($0\fday36372$ period) and $0\fmm01$ amplitude. 
The observed period and amplitude of pulsations as well as ground-based
multicolor photometry favor its classification as $\gamma$~Doradus 
type variable, however, based on our data we cannot
exclude the SPB-hypothesis for the primary.
The detailed frequency analysis suggests the presence of an additional
independent pulsation mode with the frequency of $0.21$--$0.22$~c/d.
Other detected frequencies can be explained as combinations of the above
frequencies and the orbital frequency.

The system deserves a detailed spectroscopic study which could better
constrain physical parameters of the pulsating component.
A Str{\"o}mgren photometry of the primary eclipse could also be useful for
this purpose.

\acknowledgments{Based on observations made with the WFC at the INT operated
on the island of La Palma by the Isaac Newton
Group in the Spanish Observatorio del Roque de los Muchachos of the
Instituto de Astrofisica de Canarias.
We would like to thank Sergei Antipin and Nikolai Samus
for aid in classification of this peculiar object, Suzanne Aigrain and Werner Weiss for help in
establishing this fruitful collaboration, as well as Nicola Marchili, Frank Schinzel
and the anonymous referee for reviewing this manuscript.
KS was supported through a stipend from the International Max Planck Research School (IMPRS)
for Astronomy and Astrophysics at the Universities of Bonn and Cologne.
CM and CD research has been funded by the
Italian Space Agency (ASI) under contract ASI/INAF I/015/07/00 in the
frame of the ASI-ESS project.
This research has made use of the Exo-Dat database, operated at
LAM-OAMP, Marseille, France, on behalf of the CoRoT/Exoplanet program.
The publication makes use of data products from the Two Micron All Sky
Survey, which is a joint project of the UMass/IPAC-Caltech, funded by the
NASA and the NSF, the Aladin interactive sky atlas, operated at CDS,
Strasbourg, France, the International Variable Star Index (VSX) operated
by the AAVSO and the NASA/IPAC Extragalactic Database (NED) which is operated by
the JPL, Caltech, under contract with the NASA. 
This research has made use of NASA's Astrophysics Data System.
}

\References{
\rfr Aigrain, S., Pont, F., Fressin, F., et al. 2009, A\&A, 506, 425
\rfr Antipin, S.~V., Sokolovsky, K.~V., \& Ignatieva, T.~I. 2007, MNRAS, 379L, 60
\rfr Araujo-Betancor, S., G{\"a}nsicke, B.~T., Hagen, H.-J., et al. 2005, A\&A, 430, 629
\rfr Auvergne, M., Bodin, P., Boisnard, L., et al., 2009, A\&A, 506, 411
\rfr Bessell, M.~S. \& Brett, J.~M. 1988 PASP, 100, 1134
\rfr B\'{\i}r{\'o}, I. B. \& Nuspl, J. 2005, ASPC, 333, 221
\rfr Carpano, S., Cabrera, J., Alonso, R., et al. 2009, A\&A, 506, 491 
\rfr Chochol, D., \& Pribulla, T. 2000, ASPC, 203, 125
\rfr Davidge, T.~J., \& Milone, E.~F., 1984, ApJS, 55, 571
\rfr Damiani, C., Maceroni, C., Cardini, D., et al. 2010, Ap\&SS, 66 
\rfr Debosscher, J., Sarro, L.~M., L{\'o}pez, M., et al. 2009, A\&A, 506, 519 
\rfr Deleuil, M., Moutou, C., Deeg, H.~J., et al. 2006, ESA Special Publication, 1306, 341 
\rfr Fridlund, M., Baglin, A., Lochard, J., Conroy, L. 2006, ESASP, 1306,  
\rfr Handler, G., Balona, L.~A., Shobbrook, R.~R. et al. 2002, MNRAS, 333, 262 
\rfr Hekker, S., Debosscher, J., Huber, D. et al. 2010, ApJ, 713, L187
\rfr Henry, G.~W., Fekel, F.~C., \& Henry, S.~M. 2007 AJ, 133, 1421
\rfr Ibano{\v g}lu, C., Ta{\c s}, G., Sipahi, E., \& Evren S, 2007, MNRAS, 376, 573 
\rfr Kalberla, P.~M.~W., Burton, W.~B., Hartmann, D. et al. 2005, A\&A, 440, 775
\rfr Kallinger, T., Reegen, P., Weiss, W.~W. 2008, A\&A, 481, 571 
\rfr Kaye, A.~B., Handler, G., Krisciunas, K., Poretti, E., Zerbi, F.~M. 1999, PASP, 111, 840 
\rfr Khruslov, A.~V. 2008, PZ, 28, 4
\rfr Kwee, K.~K. \& van Woerden, H. 1956, BAN, 12, 327 
\rfr Lafler, J. \& Kinman, T.~D. 1965, ApJS, 11, 216 
\rfr Liu, Q.-Y., \& Yang, Y.-L. 2003, ChJAA, 3, 142
\rfr Maceroni, C., Montalb{\'a}n, J., Michel, E. et al. 2009, A\&A, 508, 1375 
\rfr Maceroni, C. et al. 2010 in preparation; to appear in AN
\rfr Mkrtichian, D.~E., Kusakin, A.~V., Gamarova, A.~Y., \& Nazarenko, V. 2002, ASPC, 259, 96 
\rfr Mkrtichian, D.~E., Kim, S.-L., Rodr\'{\i}guez, E., et al. 2007, ASPC, 370, 194
\rfr Monet, D.~G., Levine, S.~E., Canzian, B., et al. 2003, AJ, 125, 984
\rfr Moon, T. 1986, Ap\&SS, 122, 173
\rfr Pigulski, A., \& Pojma{\'n}ski, G. 2009, AIPC, 1170, 351 
\rfr Predehl, P., \& Schmitt, J.~H.~M.~M. 1995, A\&A, 293, 889
\rfr Pr{\v s}a, A., \& Zwitter, T.\ 2005, ApJ, 628, 426 
\rfr Reegen, P. 2007, A\&A, 467, 1353
\rfr Rodr\'{\i}guez, E. 2002 ESASP, 485, 331
\rfr Samus, N.~N., Durlevich, O.~V. et al. 2009, yCat, 1, 2025
\rfr Schlegel, D.~J., Finkbeiner, D.~P., \& Davis M. 1998, ApJ, 500, 525 
\rfr Schwarzenberg-Czerny, A. 1991 MNRAS, 253, 198
\rfr Skrutskie, M.~F., Cutri, R.~M., Stiening, R. et al. 2006, AJ, 131, 1163 
\rfr Stankov, A. \& Handler, G. 2005, ApJS, 158, 193 
\rfr Thoul, A. 2009, CoAst, 159, 35
\rfr Wilson, R.~E., \& Devinney, E.~J.\ 1971, ApJ, 166, 605
}

\end{document}